\documentclass[12pt]{article}
\usepackage{amsfonts}
\usepackage{latexsym}
\usepackage{amsmath}
\usepackage{amssymb}
\usepackage{amssymb}

\newcommand{\newsection}{    
\setcounter{equation}{0}\section}
\def\appendix#1{\addtocounter{section}{1}\setcounter{equation}{0}
\renewcommand{\thesection}{\Alph{section}}
\section*{Appendix \thesection\protect\indent \parbox[t]{11.15cm}{#1}}
\addcontentsline{toc}{section}{Appendix \thesection\ \ \ #1}}

\newcommand{\be}{\begin{eqnarray}}
\newcommand{\ee}{\end{eqnarray}}
\newcommand{\bea}{\begin{eqnarray}}
\newcommand{\eea}{\end{eqnarray}}
\newcommand{\ba}{\begin{array}}
\newcommand{\ea}{\end{array}}
\newcommand{\nn}{\nonumber \\}

\def \la {\label}

\def\a{\alpha}
\def\b{\beta}

\def\e{\epsilon}

\def\bbe{{\bf{e}}}
\font\mybb=msbm10 at 11pt

\def\bb#1{\hbox{\mybb#1}}

\def\bR {\bb{R}}

\def\bH {\bb{H}}

\def\ui {{\underline {i}}}
\def\uj {{\underline {j}}}

\def\ui {{\underline {I}}}
\def\uj {{\underline {J}}}

\def\ua{{\underline {a}}}
\def\ub{{\underline {b}}}

\def\uM{{\underline {M}}}

\def\uA{{\underline {A}}}
\def\uB{{\underline {B}}}

\def\uo{{\underline {1}}}
\def\u2{{\underline {2}}}

\def\pr{{{r'}}}
\def\ps{{{s'}}}
\def\pt{{ {t'}}}

\def\ur{{\underline {r}}}
\def\us{{\underline {s}}}

\begin{document}
\begin{titlepage}
\begin{center}
\vspace{5.0cm}

\vspace{3.0cm} {\Large \bf Topology and geometry of 6-dimensional  (1,0) supergravity black hole horizons}
\\
[.2cm]

{}\vspace{2.0cm}
 {\large
M.~Akyol and  G.~Papadopoulos
 }

{}

\vspace{1.0cm}
Department of Mathematics\\
King's College London\\
Strand\\
London WC2R 2LS, UK\\

\end{center}
{}
\vskip 3.0 cm
\begin{abstract}
We show that the supersymmetric  near horizon black hole geometries of   6-dimensional supergravity coupled to any number
of scalar and tensor multiplets are either locally $AdS_3\times \Sigma^3$, where
 $\Sigma^3$ is a homology 3-sphere,  or  $\bR^{1,1}\times {\cal S}^4$,
 where ${\cal S}^4$ is a 4-manifold whose geometry depends on the hypermultiplet scalars.
 In both cases, we find that the tensorini multiplet scalars
 are constant and the associated 3-form field strengths vanish.  We also demonstrate that the $AdS_3\times \Sigma^3$ horizons
   preserve  2,  4  and 8 supersymmetries.   For horizons with 4 supersymmetries, $\Sigma^3$ is  in addition a non-trivial
   circle fibration over a topological 2-sphere.
The near horizon geometries preserving 8 supersymmetries are locally isometric to
either $AdS_3\times S^3$ or $\bR^{1,1}\times T^4$. Moreover, we show that the $\bR^{1,1}\times {\cal S}$ horizons preserve 1, 2 and 4 supersymmetries and   the geometry of ${\cal S}$ is Riemann, K\"ahler and hyper-K\"ahler, respectively.
\end{abstract}
\end{titlepage}

\setcounter{section}{0}
\setcounter{subsection}{0}


\newsection{Introduction}

It is well known that the black hole uniqueness theorems in 4-dimensions
\cite{israel}-\cite{robinson}
do not extend to 5 and higher dimensions. Specifically in 5 dimensions, apart
from spherical black holes \cite{bmpv}, there also exist black holes with near horizon topology
$S^1\times S^2$, the black rings \cite{reallbh, ring1}. In more than 5 dimensions, it is expected that there  are black holes with exotic horizon topologies \cite{gibbons1}-\cite{obers1}.

In the context of supergravity,
the recent progress made towards understanding of the geometry for all solutions to the Killing spinor equations (KSEs)  raises the
hope that all supersymmetric black hole solutions can be classified. So far this goal has not been attained but some
significant progress has been made towards classifying all near horizon black hole geometries. Results in this
direction include the identification of all near horizon geometries of
simple 5- and 6-dimensional supergravities \cite{reallbh, jgdm}. In addition,
all near horizon geometries of  4-dimensional ${\cal N}=1$ supergravity coupled
to any number of vector and scalar multiplets have been
classified  \cite{fourhor}. Furthermore,
the near horizon geometries of heterotic \cite{hh} and  IIB  with 5-form flux \cite{iibhor} supergravities have been
identified and many examples have been constructed. Progress has also been made
in 11 dimensions where all static near horizon geometries have been found \cite{mhor}.

In this paper, we shall investigate  the topology and geometry of supersymmetric black hole horizons  in 6-dimensional (1,0)
supergravity \cite{sezgin, ferrara, riccioni} coupled to any number of tensor, vector  and scalar\footnote{The scalar multiplets
are also referred to as hypermultiplets.}  multiplets. For this, we adapt null gaussian geodesic coordinates\footnote{The nature of the horizons captured by these coordinates is described in \cite{wald}. A discussion of the application
 of these coordinates to supergravity horizons with active form field strengths and the relation of the resulting solutions  to asymptotic supersymmetry algebras can be found in  \cite{hh, fourhor}.}
 near the horizon
 \cite{wald} and then  solve both the field and KSEs of the theory. The solution of the latter is  facilitated by the identification of the geometry of all supersymmetric backgrounds
of 6-dimensional (1,0) supergravity in \cite{akyolgp}. There is not a general method to solve the field equations. However
in many cases of interest, the use of the results from the KSEs together with  the maximal principle
and the compactness of the
horizon sections allow for the general solution to the field equations without imposing an ansatz  on the form of near horizon geometries.
To apply this technique in 6 dimensions, we shall restrict our attention
 to those  horizons for which the vector multiplets can be consistently set to zero. This is because in the presence of active vectors  the field equations of the theory cannot be put in a form that allows the application of the maximal principle\footnote{This is similar to what happens
in the heterotic case and a more detailed explanation is given in \cite{hh}.}.

In particular, we find that there
are two classes of near horizon geometries.
The near horizon geometries of the first
class are locally $AdS_3\times \Sigma^3$ and so the horizon section is ${\cal S}=S^1\times \Sigma^3$.  The proof of this product structure for ${\cal S}$ is key and utilizes the field and KSEs as well as the compactness of ${\cal S}$. Moreover, it turns out that $\Sigma^3$ is a 3-dimensional manifold
with strictly positive Ricci tensor and  a theorem of Gallot and Meyer, see eg \cite{ppetersen},  together
with the Poincar\'e conjecture \cite{poincare} imply that the universal cover of $\Sigma^3$ is diffeomorphic to $S^3$.
The scalars of the tensor multiplets are constant
and the associated 3-form field strengths vanish. The scalars of the hypermultiplets
depend only on the coordinates of $\Sigma^3$ and  satisfy a natural first
order non-linear differential equation which we describe.

We  also demonstrate that the $AdS_3\times \Sigma^3$
horizons preserve 2, 4 or 8 supersymmetries depending on the geometry of $\Sigma^3$. To prove  this, we show that the horizons admit an isometry which commutes with all the KSEs.
For a generic choice of $\Sigma^3$, the horizons preserve 2 supersymmetries. For horizons preserving 4 supersymmetries,
the metric on $\Sigma^3$ is compatible with a non-trivial  circle fibration over a topological 2-sphere $\Sigma^2$.
Moreover the scalars of the hypermultiplet depend only on the coordinates
of $\Sigma^2$ and are pseudo-holomorphic type of maps into the hypermultiplets Quaternionic K\"ahler
manifold ${\cal Q}$. The $AdS_3\times \Sigma^3$ horizons which preserve 8 supersymmetries
are locally isometric\footnote{Throughout this paper,  $S^n$ denotes the n-sphere equipped with the
 standard ``round'' metric.} to $AdS_3\times S^3$.

 The geometry of the horizon sections ${\cal S}$ in the $\bR^{1,1}\times {\cal S}$ class of horizons depends only on the hypermultiplet scalars. In particular as in the previous case,
 the tensor multiplet scalars are constant and all 3-form field strengths, including that of the gravity multiplet,  vanish.
The $\bR^{1,1}\times {\cal S}$ horizons preserve 1, 2 and 4 superymmetries, and the geometry
of ${\cal S}$ is Riemann, K\"ahler and hyper-K\"ahler, respectively. The first two cases require the existence of non-trivial hypermultiplet scalars which satisfy certain first
order non-linear differental equations which we describe. In the last case, the hypermultiplet
scalars are constant and ${\cal S}$ is locally isometric to either $K_3$ or $T^4$. The
$\bR^{1,1}\times T^4$ horizons admit a supersymmetry enhancement to $N=8$. The above results  extend those of \cite{jgdm} on the near horizon geometries of simple 6-dimensional supergravity.

This paper has been organized as follows. In section 2, we set up our notation and solve the KSEs for $N=1$ horizons. In section 3, we  prove that all near horizon
geometries with non-vanishing ``rotation'' are products $AdS_3\times  {\Sigma^3}$ and preserve even number of supersymmetries. Moreover, we examine in detail
the topology and geometry of $N=2$ horizons. In section 4, we investigate the geometry of $AdS_3\times {\Sigma^3}$ horizons which preserve 4 and 8 supersymmetries.  In section 5,  we examine the geometry of $\bR^{1,1}\times{\cal S}$
horizons and show that they may preserve 1,2 and 4 supersymmetries.
In appendix A, we present the solution of the KSEs for near horizon backgrounds preserving one supersymmetry.

\newsection{Supersymmetric horizons }

\subsection{Fields and KSEs of 6-dimensional supergravity}

We consider a 6-dimensional supergravity coupled to any number of tensor and scalar multiplets.
 The bosonic field content of the gravitational multiplet consists of the graviton and 3-form field strength $H$ of the gravitational multiplet. In addition, each tensor
 multiplet contains  a 3-form field strength and a real scalar, while each scalar multiplet,
 or hyper-multiplet,
   contains 4 real scalars.   The action of the theory
   has been constructed progressively in \cite{sezgin, ferrara, riccioni}.  We follow closely the description of the theory in \cite{riccioni}
    but we shall use the notation developed in \cite{akyolgp} which  differs from that in \cite{riccioni}. Apart from differences in the normalizations of fields, the 10-dimensional description  of (1,0)-supergravity spinors in \cite{akyolgp} simplifies
   the solution of the KSEs.

 Suppose that the (1,0)-supergravity couples to $n_T$ tensor multiplets. In the absence of vector multiplets, the $n_T+1$ 3-form field strengths of the supergravity theory are given in terms of 2-form
 potentials as
   \be
G^\ur_{\mu\nu\rho}=3\partial_{[\mu}B^{\ur}_{\nu\rho]}~,~~~r=0,\dots, n_T~.
\ee
The gravitini, tensorini and hyperini  KSEs of the theory are
\bea
\mathcal{D}_\mu \epsilon \equiv \big(\nabla_\mu - \frac{1}{8}H_{\mu\nu\rho}\Gamma^{\nu\rho}+\mathcal{C}_\mu^{r'}\rho_{r'}\big)\e &=&0~,
\cr
\bigg(\frac{i}{2}T_\mu^{\underline{M}}\Gamma^{\mu}-\frac{i}{24}H^{\underline M}_{\mu\nu\rho}\Gamma^{\mu\nu\rho}\bigg)\epsilon &=&0~,
\cr
i\Gamma^{\mu}\epsilon_{\underline{A}}V_\mu^{\underline{aA}}&=&0~,
\la{kses}
\eea
respectively, where the various fields and components of the KSEs are defined as
\be
H_{\mu\nu\rho} & = & v_{\underline{r}}G_{\mu\nu\rho}^{\underline{r}},~~~ H_{\mu\nu\rho}^{\underline{M}}=x_{\underline{r}}^{\underline{M}}G_{\mu\nu\rho}^{\underline{r}}~,~~~{\mathcal{C}_{\mu}}^{\underline{A}}{}_{\underline{B}} =\partial_{\mu}\phi^{\underline{I}}{\mathcal{A}_{\underline{I}}}^{\underline{A}}{}_{\underline{B}}~,\nonumber \\
\cr
T_{\mu}^{\underline{M}} & = & x_{\underline{r}}^{\underline{M}}\partial_{\mu}v^{\underline{r}}~,~~~ V_{\mu}^{\underline{aA}}=E_{\underline{I}}^{\underline{aA}}\partial_{\mu}
\phi^{\underline{I}}~,
\label{Fielddef}
\ee
and where
\bea
\eta_{\ur\us} v^\ur v^\us=1~,~~~v_\ur v_\us- \sum_\uM x_\ur^\uM x_\us^\uM=\eta_{\ur\us}~,~~~v^\ur x_\ur^\uM=0~.
\eea
Both $v_\ur$ and $x_\us^\uM$ depend on the tensor multiplet scalars $\varphi$.
For more notation details as well as  the description of spinors, see \cite{akyolgp}.
Clearly $H^{\underline{M}}$ are the $n_T$ 3-form field strengths of the tensor multiplets, and the tensor multiplet scalars $\varphi$ parameterize the hyperbolic space $SO(n_T,1)/SO(n_T)$. $\phi^{\underline{I}}$ are the scalars of the hyper-multiplets which
take values on a Quaternionic K\"ahler manifold, ${\cal Q}$, $E_{\underline{I}}^{\underline{aA}}$
is a frame of ${\cal Q}$, and ${\mathcal{A}_{\underline{I}}}^{\underline{A}}{}_{\underline{B}}$ is the $Sp(1)$ part of the quaternionic K\"ahler connection. Note also that
the 3-form field strengths satisfy the duality condition
\bea
\zeta_{\ur\us} G^\us_{\mu_1\mu_2\mu_3}={1\over 3!} \epsilon_{\mu_1\mu_2\mu_3}{}^{\nu_1\nu_2\nu_3} G_{\ur \nu_1\nu_2\nu_3}~,
\eea
where
\bea
\zeta_{\ur\us}= v_\ur v_\us+ \sum_\uM x^\uM_\ur x^\uM_\us~,
\eea
ie $H$ is anti-self-dual while $H^\uM$
is self-dual.

The KSEs of (1,0)-supergravity, including that of the vector multiplet which
 has not been given above, have been solved for all backgrounds preserving any number
 of supersymmetries in \cite{akyolgp}. Here, we shall adapt the analysis to describe
 the topology and geometry of all supersymmetric  horizons.

 \subsection{Near horizon geometry}

The description of the fields of (1,0) supergravity  near the  horizon
of an extreme black hole using null Gaussian coordinates \cite{wald} is similar to that which has been given for the fields of  heterotic
supergravity  in \cite{hh}. Because of this, we shall not present the details here. In particular, the near horizon  metric, 3-form field strengths and scalars of (1,0) supergravity  can be written as
\bea
ds^2&=&2 \bbe^+ \bbe^- + \delta_{ij} \bbe^i \bbe^j~,
\cr
G^\ur &=&  \bbe^+ \wedge \bbe^- \wedge \big(dS^\ur-N^\ur-S^\ur h\big)
+ r \bbe^+ \wedge \big(h \wedge N^\ur - dN^\ur - S^\ur dh \big) + dW^\ur~,
\cr
\phi^{\underline{I}}&=&\phi^{\underline{I}}(y)~,~~~~\varphi= \varphi(y)~,
\la{bhdata}
\eea
where
\bea
\bbe^+ = du~,~~~
\bbe^- = dr + r h+r^2\Delta du ~,~~~
\bbe^i &=& e^i{}_I dy^I~.
\label{nhbasis}
\eea
 The spacetime has coordinates $(r, u, y^I)$. The black hole horizon section ${\cal S}$ is the co-dimension 2 subspace $r=u=0$ and it is assumed to be {\it compact}, {\it connected}, and {\it without boundary}. The dependence of fields  on light-cone coordinates $(r,u)$ is explicitly given. In addition, $W^\ur$ are  2-forms,  $h, N^\ur$ are 1-forms, and $S^\ur$ are scalars on the horizon section ${\cal S}$ and  depend only on the coordinates $y^I$.
$\bbe^i$ is a frame on ${\cal S}$ and depends only on $y$ as well. Both the tensor and  hypermultiplet scalars depend only on the coordinates of ${\cal S}$.

To find the supersymmetric horizons of 6-dimensional (1,0) supergravity, one has to solve both the field and KSEs of the theory for the fields given in (\ref{bhdata}).
We shall proceed with the solution of KSEs.

\subsection{Solution of KSEs}

To continue, we substitute (\ref{bhdata}) into the KSEs (\ref{kses}) and assume that
the backgrounds preserve at least one supersymmetry.  Furthermore, we identify the stationary Killing vector field $\partial_u$ of the near horizon geometry with the Killing vector constructed as a Killing spinor bilinear. A detailed analysis of these calculations has been presented
 in appendix A. The end result of this computation  is that the Killing spinor can be chosen as
\bea
\e=1+e_{1234}~,
\eea
and the fields can be rewritten as
\bea
 ds^2&=& 2 \bbe^+ \bbe^-+ \delta_{ij} \bbe^i \bbe^j~,
\cr
H&=&{\bf e}^+ \wedge {\bf e}^- \wedge h+r{\bf e}^+ \wedge dh-{1\over 3!} h_{\ell}\,\,\epsilon^\ell{}_{ijk} \,\, \bbe^i\wedge \bbe^j\wedge \bbe^k~~,
\cr
H^\uM&=& T_i^{\underline M}\,\,{\bf e}^- \wedge {\bf e}^+ \wedge {\bf e}^i- {1\over3!}
T^\uM_\ell\,\epsilon^\ell{}_{ijk}\,\,\bbe^i\wedge \bbe^j\wedge \bbe^k ~.
\cr
\phi^{\ui}&=&\phi^\ui(y)~,~~~\varphi=\varphi(y)~,
\la{n1fields}
\eea
where we have used the duality relations of the 3-form field strengths. In
particular $h_{\ell}\,\,\epsilon^\ell{}_{ijk}= -v_{\underline r} dW^{\underline r}_{ijk}$ and similarly
for $H^\uM$. In addition the anti-self duality of $H$ requires that
\bea
dh_{ij}=-{1\over2} \e_{ij}{}^{kl} dh_{kl}~.
\eea
It is clear that $H$ is entirely determined in terms of $h$ while
$H^\uM$ is entirely determined in terms of the scalars $\varphi$ of the tensor
multiplets.

Furthermore, the gravitino KSE along the horizon section directions
requires that
\bea
\tilde{\cal D}_i (1+e_{1234})=0~,
\la{sgkse}
\eea
where
\bea
\tilde{\cal D}_i=\hat{\tilde \nabla}_i+\mathcal{C}^{r'}_i\rho_{r'}~,
\eea
and $\hat{\tilde \nabla}$ is the connection on ${\cal S}$ with skew-symmetric
torsion $-\star_4 h$. One can unveil the geometric content of this equation by
considering the Quaternionic-Hermitian 2-forms
\bea
\omega_I=-i\delta_{\a\bar\b} e^\a\wedge e^{\bar\b}~,~~~\omega_J=-e^1\wedge e^2-e^{\bar 1}\wedge e^{\bar 2}~,~~~
\omega_K=i(e^1\wedge e^2-e^{\bar 1}\wedge e^{\bar 2})~.
\la{qhforms}
\eea
on ${\cal S}$ which can be constructed as twisted Killing spinor bi-linears, see \cite{akyolgp}. In particular setting  $\omega^1=\omega_I$, $\omega^2=\omega_J$ and $\omega^3=\omega_K$, the integrability condition of (\ref{sgkse}) can be expressed as

\bea
-\hat {\tilde R}_{mn,}{}^k{}_i \omega^{\pr}{}_{kj}+(j,i)+2{\cal F}^{\ps}_{mn}
\epsilon^{\pr}{}_{\ps\pt} \omega^{\pt}_{ij}=0~,
\la{1int1}
\eea
where
\bea
{\cal F}^{\ps}_{mn}=\partial_m \phi^\ui \partial_n \phi^\uj {\cal F}^\ps_{\ui\uj}~.
\eea
The integrability condition identifies the   $Sp(1)\subset Sp(1)\cdot Sp(1)$ component of the
curvature $\hat {\tilde R}$ of the 4-dimensional  manifold  ${\cal S}$ with the pull back with respect to $\phi$ of the $Sp(1)$ component of the curvature of  ${\cal Q}$.
The restriction imposed on the geometry of ${\cal S}$ by (\ref{1int1})
depends on the scalars $\phi^\ui$. In particular, if $\phi^\ui$ are constant, then
${\cal F}_{mn}=0$ and
(\ref{1int1})  implies that ${\cal S}$ is an HKT manifold \cite{hkt}.

There are no additional conditions arising from the tensorini KSE. The hyperini KSE
implies in addition that
\bea
-V_1^{\ua\uo}+ V_{\bar2}^{\ua\u2}=0~,~~~ V_2^{\ua\uo}+ V_{\bar1}^{\ua\u2}=0~.
\la{1hyp1}
\eea
We shall return on all the above conditions imposed by the KSEs after  imposing the restrictions on the fields implied
by the field equations of the theory and the compactness of ${\cal S}$.

\newsection{Horizons with $h\not=0$}

There are two classes of horizons to consider depending on whether or not $h$ vanishes.
First, we shall consider the case that $h\not=0$.

\subsection{Holonomy reduction}

If $h\not=0$, we shall demonstrate that, as in the heterotic case,  the number of supersymmetries preserved by the near horizon geometries is always even. For this we shall use the results we have obtained
 from the KSEs for
horizons preserving one supersymmetry and the field equations of the theory.
The methodology we shall  follow to prove this is to compute $\tilde\nabla^2 h^2$ and
apply the maximum principle utilizing the compactness of ${\cal S}$.

The field  equations of 6-dimensional supergravity in the absence of vector multiplets are
\bea
R_{\mu \nu}- \frac{1}{4} \varsigma_{\ur\us}{G^\ur_{\mu}}^{\alpha\beta}G^\us_{\nu\alpha\beta} + \partial_{\mu}v^\ur\partial_{\nu}v_\ur
- 2g_{\ui\uj}\partial_{\mu}\phi^\ui\partial_{\nu}\phi^\uj&=&0~,\cr
\nabla_{\lambda}\big(\varsigma_{\ur\us}G^{\us\lambda\mu\nu}\big)&=&0~,\cr
\nabla^{\mu}\partial_{\mu}v^\ur + \frac{1}{6}v_\us G^{\us\mu\nu\rho}G^\ur_{\mu\nu\rho}&=&0~,\cr
D_{\mu}\partial^{\mu}\phi^\ui &=&0~,
\la{feqns}
\eea
where in the last equation it is understood that the Levi-Civita connections
of both the spacetime and the hyper-multiplets Quaternionic K\"ahler manifold metrics have been used to covariantize the expression.

First one finds that
\bea
\tilde{\nabla}^2 h^2 = 2\tilde{\nabla}^ih^j\tilde{\nabla}_ih_j + 2\tilde{\nabla}^i(dh)_{ij}h^j + 2\tilde{R}_{ij}h^ih^j + 2h^j\tilde{\nabla}_j\tilde{\nabla}_ih^i~,
\label{h2}
\eea
where $\tilde \nabla$ is the Levi-Civita connection of ${\cal S}$ with respect
to $ds^2({\cal S})=\delta_{ij} \bbe^i \bbe^j$ and $\tilde R$ is the associated Ricci tensor.
The proof of this is given in \cite{hh}.
To proceed, we shall utilize  the field equations to rearrange the above expression
in such a way that we can apply the maximum principle. Using  the Einstein equation
and
\bea
\tilde{R}_{ij}= R_{ij}-\tilde{\nabla}_{(i}h_{j)} + \frac{1}{2}h_ih_j~,
\eea
one finds that
\bea
2\tilde{R}_{ij}h^ih^j&=&-h^2{\partial}_kv_\ur{\partial}^kv^\ur  + 4{\partial}_i\phi^\ui{\partial}_j\phi^\uj g_{\ui\uj}h^ih^j - h^i\tilde{\nabla}_ih^2~.
\label{rij}
\eea
The $\mu\nu=+-$  component of the field equation $\nabla_{\lambda}\big(\varsigma_{\ur\us}G^{\us\lambda\mu\nu}\big)$
together with $H^{i+-}=-h^i$ and $H^{{\underline{M}}i+-}=T^{i{\underline{M}}}$ give
\bea
{\partial}_iv_\ur h^i+v_\ur\tilde{\nabla}_ih^i+\tilde{\nabla}_i{\partial}^iv_\ur=0~.
\la{gmn}
\eea
Acting on the above expression with $v^\ur$, we find
\bea
\tilde{\nabla}_ih^i+v^\ur\tilde{\nabla}_i {\partial}^iv_\ur=0~,
\la{xx1}
\eea
where we have used $v_\ur v^\ur=1$.

The field equation of the scalars of the tensor multiplet gives
\bea
v_\ur\tilde{\nabla}_i{\partial}^iv^\ur =0~,
\la{xx2}
\eea
which when combined with (\ref{xx1})  implies that
\bea
\tilde{\nabla}_ih^i=0~.
\la{xx4}
\eea
In addition (\ref{xx2}) and  $v_\ur v^\ur=1$ give
\bea
{\partial}_kv_\ur{\partial}^kv^\ur=0~.
\la{xx3}
\eea
Thus substituting (\ref{rij}) into (\ref{h2}) and using (\ref{xx4}) and (\ref{xx3}),  we find that
\bea
\tilde{\nabla}^2 h^2+h^i\tilde{\nabla}_ih^2 = 2\tilde{\nabla}^ih^j\tilde{\nabla}_ih_j + 2\tilde{\nabla}^i(dh)_{ij}h^j + 4{\partial}_i\phi^\ui{\partial}_j\phi^\uj g_{\ui\uj}h^ih^j~.
\label{h3}
\eea
This expression is close to the one required for the maximum principle to apply. It remains to determine $dh$. For this, consider the $jk$-component of the 3-form  field equation to find
\bea
\nabla^i(v_\ur H_{ijk}+x_\ur^{\underline M}H^{\underline M}_{ijk})=\epsilon_{ijkl} \partial^iv_\ur h^l+
v_\ur\epsilon_{ijkl}\nabla^ih^l =0~,
\eea
which implies that
\bea
dh =0~,
\eea
Substituting this into  (\ref{h3}), we get
\bea
\tilde{\nabla}^2 h^2+h^i\tilde{\nabla}_ih^2 = 2\tilde{\nabla}^ih^j\tilde{\nabla}_ih_j  + 4{\partial}_i\phi^\ui{\partial}_j\phi^\uj g_{\ui\uj}h^ih^j~.
\label{h4}
\eea
Applying now the maximum principle using  the compactness of ${\cal S}$, we find
  that $h^2$ is constant and
\bea
\tilde{\nabla}_ih_j&=&0~,
\cr
h^i{\partial}_i\phi^\ui&=&0~.
\la{hphi}
\eea
To establish the latter equation, we have used that the metric of the hyper-multiplets Quaternionic
K\"ahler manifold is positive definite.
Thus $h$ is a parallel 1-form on ${\cal S}$ with respect to the Levi-Civita connection
and the scalars of the hyper-multiplets are invariant under the action of $h$. Note also
that $\hat{\tilde{\nabla}}h=0$ as $i_h \tilde H=0$.

The existence of a parallel 1-form on the horizon section ${\cal S}$ with respect
to the Levi-Civita connection is a strong restriction. First it implies that
the holonomy of $\tilde \nabla$ is contained in $SO(3)\subset SO(4)$,
\bea
{\rm hol} (\tilde \nabla)\subseteq SO(3)~.
\eea
 Moreover
${\cal S}$ metrically  (locally) splits into a product $S^1\times \Sigma^3$, where $\Sigma^3$ is a 3-dimensional manifold. In turn, as we shall see, the near horizon
 geometry is locally a product $AdS_3\times \Sigma^3$. More elegantly the near horizon geometry admits a supersymmetry enhancement from one supersymmetry to two.

\subsection{Supersymmetry enhancement}

To demonstrate supersymmetry enhancement for the backgrounds  with $h\not=0$,
let us re-investigate the KSEs for the fields given in (\ref{n1fields}). It is straightforward to see by substituting (\ref{n1fields}) into the KSEs and following
the calculation in appendix A that the general form of a Killing spinor is
\bea
\epsilon=\eta_+-{u\over2} h_i \Gamma^i\Gamma_+ \eta_-+ \eta_-
\eea
where $\eta_\pm$ depend only on the coordinates of ${\cal S}$.  In addition
the gravitino KSE requires that
\bea
\hat{\tilde \nabla}_i\epsilon+ \mathcal{C}^{r'}_i\rho_{r'}\epsilon=0~,
\la{grav2}
\eea
the tensorini KSE implies that
\bea
(1\pm {1\over2}) T_i^\uM\Gamma^i\epsilon_\pm-{1\over12} H^\uM_{ijk} \Gamma^{ijk}\epsilon_\pm=0~,
\eea
and the hyperini KSE gives
\bea
i\Gamma^i\epsilon_{\pm\underline{A}}V_i^{\underline{aA}}=0~.
\la{hyper2}
\eea
Next we shall show that both
\bea
\epsilon_1=1+e_{1234}~,~~~\epsilon_2= \Gamma_- h_i \Gamma^i (1+e_{1234})-u k^2 (1+e_{1234})~,
\la{ksp2}
\eea
are Killing spinors, where we have set $k^2=h^2$ for the constant length of $h$. Observe that the second Killing spinor is constructed by setting
$\eta_+=0$ and $\eta_-=\Gamma_- h_i \Gamma^i (1+e_{1234})$.

We have already
solved the KSEs for $\epsilon_1$. Next observe that $\epsilon_2$ solves
the gravitino KSE as the Clifford algebra operation $h_i\Gamma^i \Gamma_-$ commutes
with the supercovariant derivative in (\ref{grav2}) as a consequence of the
reduction of holonomy demonstrated in the previous section. In addition,
the same Clifford operation commutes with the hyperini KSE  as a result of the second eqn in (\ref{hphi}) and (\ref{hyper2}).

It remains to show that $\epsilon_2$ solves the tensorini KSE as well. This
is a consequence of (\ref{xx3}). For this observe that the metric induced on
 $SO(n_T, 1)/SO(n_T)$  by the algebraic equation  $\eta_{\ur\us} v^\ur v^\us=1$ is the
 standard hyperbolic metric. So it has Euclidean signature. As a result,
\bea
\partial_i v^\ur=0~.
\la{xu1}
\eea
Thus, we conclude that the scalar fields are constant and the 3-form
field strengths of the tensorini multiplet vanish.
This  agrees with the classification results of \cite{akyolgp} for solutions
of the KSEs of 6-dimensional supergravity preserving at least two supersymmetries
whose Killing spinors have   compact isotropy  group. Some of the results of this section are tabulated in table 1.

\begin{table}
\centering
\fontencoding{OML}\fontfamily{cmm}\fontseries{m}\fontshape{it}\selectfont
\begin{tabular}{|c|c|c|c|}\hline
${\rm Iso}(\eta_+)$& ${\rm hol}({\tilde{\cal D}})$ &$N$&$ \eta_+$
 \\
\hline\hline
  $Sp(1)\cdot Sp(1)\ltimes\bH$ & $Sp(1)$ & $2$ & $1+e_{1234}$
 \\
 \hline
$U(1)\cdot Sp(1)\ltimes\bH$&$U(1)$&$4$&$1+e_{1234}~, ~i(1-e_{1234})$
\\
\hline
$Sp(1)\ltimes \bH^4$&$\{1\}$&$8$&$1+e_{1234}~, ~i(1-e_{1234})~,~e_{12}-e_{34}~,~i(e_{12}+e_{34})$
\\
\hline
\end{tabular}
\label{tab1}
\begin{caption}
{\small {\rm ~~Some of the geometric data used to solving the gravitino KSE are described.
In the first column, we give the isotropy groups, ${\rm Iso}(\eta_+)$, of $\{\eta_+\}$ spinors in $Spin(5,1)\cdot Sp(1)$. In the second column
we state the holonomy of the supercovariant connection $\tilde{{\cal D}}$ of the horizon section ${\cal S}$ in each case. The holonomy of $\hat{\tilde\nabla}$ is identical to that of $\hat\nabla$.
In the third column, we present the number of ${\cal D}$-parallel spinors
and in the last column we give representatives of the $\{\eta_+\}$ spinors.
}}
\end{caption}
\end{table}

\subsection{Geometry}

To investigate the geometry of spacetime, one can compute the
form  bi-linears associated with the Killing spinors (\ref{ksp2}). In particular, one
finds that the spacetime admits 3 $\hat\nabla$-parallel 1-forms given by
\bea
\lambda^- = {\bf e}^-~,~~~\lambda^+ = {\bf e}^+ - \frac{1}{2}k^2u^2{\bf e}^- - uh~,~~~\lambda^1 = k^{-1}(h+k^2u{\bf e}^-)~.
\la{n21forms}
\eea
Moreover, the Lie algebra of the associated vector fields closes in $\mathfrak{sl}(2, \bR)$. To verify this, see \cite{hh}. Since $h$ is $\tilde\nabla$-parallel,  the spacetime is locally metrically a product $
SL(2, \bR)\times \Sigma^3$, ie
\bea
ds^2&=&ds^2(SL(2, \bR))+ ds^2(\Sigma^3)~,
\cr
H&=&d{\rm vol}(SL(2, \bR))+ d{\rm vol}(\Sigma^3)~,
\cr
\phi^\ui&=&\phi^\ui(z)~,
\eea
where the scalars of the hyper-multiplet depend only on the coordinates $z$ of
$\Sigma^3$.

In addition to the 1-forms given in (\ref{n21forms}), the spacetime admits
3 more twisted 1-forms bilinears, see \cite{akyolgp}. For the Killing spinors  (\ref{ksp2}), these are given by
\bea
e^{r'}= k^{-1} h_j (J^{r'})^j{}_i \bbe^i~,
\la{ers}
\eea
where $J^{r'}$ is a quaternionic structure on ${\cal S}$  associated with the  Quaternionic-Hermitian 2-forms (\ref{qhforms}).
As it has been already mentioned, these Quaternionic-Hermitian 2-forms are constructed from twisted spinor
bi-linears and so rotate to each other under patching conditions.
Observe that the frame $e^{r'}$ is orthogonal to $h$ and the rotation between the
$\bbe^i$ and $(h, e^{r'})$ is in $SO(4)$. Therefore $(k^{-1}h, e^{r'})$ is another frame
on ${\cal S}$ with $e^{r'}$ adapted to $\Sigma^3$. Thus
$ds^2({\cal S})=k^{-2} h^2+ds^2(\Sigma^3)$ with  $ds^2(\Sigma^3)=\delta_{r's'}
e^{r'} e^{s'}$.

The metric on $\Sigma^3$ is restricted by  the Einstein equation (\ref{feqns})
and the integrability condition (\ref{1int1}). The former gives
\bea
R^{(3)}_{r's'}-{1\over2} k^2 \delta_{r's'}-2 \partial_{r'} \phi^\ui \partial_{s'} \phi^\uj g_{\ui\uj}=0~,
\la{sigma3}
\eea
where $r', s'$ are indices of $\Sigma^3$ and $R^{(3)}$ is the Ricci tensor of $\Sigma^3$.
This is an equation which determines the metric on $\Sigma^3$ in terms of  $h$
and the hyper-multiplet scalars $\phi$. The integrability condition (\ref{1int1})
does not give an independent condition on the metric of $\Sigma^3$.

It remains to find the restriction imposed by supersymmetry on the scalars
$\phi$ of the hyper-multiplet. As we have shown these depend only on the coordinates
of $\Sigma^3$. A direct observation reveals that after an appropriate identification of
the frame directions of ${\cal S}$ with the Pauli matrices  $ \sigma_{s'}$, the supersymmetry conditions
can be rewritten as
\bea
\partial_{r'} \phi^\ui=-\epsilon_{r'}{}^{s't'}\, (I_{s'})^\ui{}_\uj\, \partial_{t'}\phi^\uj~,
\la{hypereqn2}
\eea
where we have used that $(I_{s'})^{\uA\ua}{}_{\uB\ub}=-i\, \sigma_{s'}{}^\uA{}_\uB \delta^\ua{}_\ub$.
This is a rather natural condition constraining the maps $\phi$ from
$\Sigma^3$ into ${\cal Q}$.
Constant maps are solutions.

The geometry on $\Sigma^3$ is determined by (\ref{sigma3}) and depends on the solutions of (\ref{hypereqn2}). For the constant map solutions of (\ref{hypereqn2}), $\Sigma^3$
is locally isometric to $S^3$ equipped with the round metric,  and so the near horizon geometry is $AdS_3\times S^3$.

Next suppose  the existence of non-trivial solutions for the equation (\ref{hypereqn2}), and upon substitution the existence of solutions for  (\ref{sigma3}). An priori one expects that the geometry on $\Sigma^3$
 depends on the choice of quaternionic K\"ahler manifold ${\cal Q}$
 for the hyper-multiplets and the choice of a solution of (\ref{hypereqn2}). However, the differential structure on $\Sigma^3$
 is independent of these choices. To show this first observe that
the Ricci tensor $R^{(3)}$ is strictly positive.  This turns out to be sufficient to determine the topology
 on $\Sigma^3$. To see this note that in 3 dimensions the Ricci
tensor determines the curvature of a manifold. Next, the strict positivity of the
Ricci tensor implies that the (reduced) holonomy of the Levi-Civita connection of $\Sigma^3$ is $SO(3)$. Then a result of Gallot and Meyer, see \cite{ppetersen}, implies that $\Sigma^3$ is a homology 3-sphere. A brief proof of this is as follows. Since
the holonomy of the Levi-Civita connection of $\Sigma^3$ is $SO(3)$, the only parallel
forms are the constant real maps and the volume form of the manifold. On the other
 hand, the positivity of the Riemann curvature tensor implies that all harmonic forms are parallel and the fundamental group is finite. Thus de Rham cohomology of $\Sigma^3$ coincides with that of $S^3$ and so $\Sigma^3$ is a homology 3-sphere. In addition since the fundamental group is finite, the universal cover of $\Sigma^3$ is compact and so by the Poincar\'e conjecture \cite{poincare}  homeomorhic, and so diffeomorphic\footnote{There is a unique differential structure on the topological 3-sphere.}, to  the 3-sphere.
 The above result implies that in the simply connected case and for non-constant solutions to (\ref{hypereqn2}), the geometry of the round sphere is deformed in such a way that  the differential, and so topological, structure of $S^3$ is maintained.

The existence of non-trivial solutions to  (\ref{hypereqn2}) is an open problem which may depend on the choice of
quaternionic K\"ahler manifold ${\cal Q}$ of the hypermultiplets. However, as we shall see horizons that preserve 8 supersymmetries
require  $\phi$ to be constant. This is compatible with the assertion made in the attractor mechanism, see
\cite{ferrara2} for the 6-dimensional supergravity case, that all the scalars take constant values at the horizon.
However, it is worth noting that the field and KSEs do not a priori imply that the scalars are constant
for near horizon geometries which preserve a small number of supersymmetries. For this some further investigation
is required which may be case dependent.

\newsection{N=4 and N=8 horizons}

\subsection{N=4 horizons}

We have shown that if $h\not=0$, the near horizon geometries preserve 2, 4 or 8
supersymmetries. We have already investigated the case with 2 supersymmetries. The
two additional Killing spinors of horizons with 4 supersymmetries can be chosen as
\bea
\epsilon^3 = i(1-e_{1234})~,~~~\epsilon^4 = -ik^2u(1-e_{1234}) + ih_i\Gamma^{+i}(1-e_{1234})~.
\eea
Observe that $\epsilon^3=\rho^1\epsilon^1$ and $\epsilon^4=\rho^1\epsilon^2$. Thus
the KSEs must commute with $\rho^1$.  As a result $\omega_1$ is a well-defined Hermitian form on ${\cal S}$. The 1-form $\hat\nabla$-parallel spinor bilinears are
\bea
\lambda^- &=& {\bf e}^-~,~~~\lambda^+ = {\bf e}^+ - \frac{1}{2}k^2u^2{\bf e}^- - uh~,~~~\lambda^1 = k^{-1}(h+k^2u{\bf e}^-)~,\nn
\lambda^4 &=& e^1~,
\label{n4bilinears}
\eea
where the first 3 bilinears are those of horizons with two supersymmetries and
$e^1$ is given in (\ref{ers}).
The associated vector fields are Killing and their Lie algebra is $\mathfrak{sl}(2, \bR)\oplus \mathfrak{u}(1)$.

The spacetime is locally metrically a product $AdS_3\times \Sigma^3$, as for
horizons preserving 2 supersymmetries. In addition in this case, $\Sigma^3$ is
a $S^1$ fibration over a 2-dimensional manifold $\Sigma^2$. The fibre direction
is spanned by $\lambda^4=e^1$. Thus
\bea
ds^2(\Sigma^3)=(e^1)^2+ ds^2(\Sigma^2)~,~~~ds^2({\cal S})=k^{-2} h^2+(e^1)^2+ ds^2(\Sigma^2)~.
\eea
Observe that $de^1\not=0$ as $e^1\wedge de^1$ is proportional to $\tilde H= d{\rm vol}(\Sigma^3)$,  and so the fibration is  twisted.

 It remains
to specify the topology of $\Sigma^2$. For this first observe that from the
results of \cite{akyolgp}, the hyper-multiplet scalars depend only on the coordinates
of $\Sigma^2$. To specify the topology of $\Sigma^2$, we compute the Ricci tensor $R^{(2)}$
of $\Sigma^2$ using the Einstein equation and in particular (\ref{sigma3}) to find
\bea
R^{(2)}_{r's'}-{1\over2} de^1_{t'u'} (de^1)^{t'u'} \delta_{r's'}-{1\over2} k^2 \delta_{r's'}-2 \partial_{r'} \phi^\ui \partial_{s'} \phi^\uj g_{\ui\uj}=0~,
\la{sigma2}
\eea
where now $r',s',t',u'$ are indices in $\Sigma^2$. It is clear that the Ricci tensor
of $\Sigma^2$ is strictly  positive and so $\Sigma^2$ is topologically a sphere irrespective
of the properties of the maps $\phi$.

We have already mentioned that the hyper-multiplet scalars $\phi$ depend only
on the coordinates of $\Sigma^2$ as a consequence of the hyperini KSE. Thus
they are maps from $\Sigma^2$ into the Quaternionic K\"ahler manifold
of the hyper-multiplets. In addition the hyperini KSE implies that
\bea
V_2^{\ua \underline 1}=0~,~~~V_{\bar{2}}^{\ua\underline2}=0~,
\la{hypereqn4}
\eea
which is equivalent to  (\ref{hypereqn2}) after additionally requiring that the scalars do not depend on the fibre direction $\lambda^4$. These conditions imply that
$\phi$ are pseudo-holomorphic maps from $\Sigma^2$ into the Quaternionic K\"ahler
manifold of the hyper-multiplets. The analysis we have made for the existence of non-constant solutions to (\ref{hypereqn2})
applies to (\ref{hypereqn4}) as well.

\subsection{N=8 horizons}

As in the cases with 2 and 4 supersymmetries, one can show that the spacetime
is locally $AdS_3\times \Sigma^3$. In addition for horizons with 8 supersymmetries,
 the  hyperini KSE implies  that the scalars of the hyper-multiplet are constant \cite{akyolgp}. In such case, the Einstein equation implies that $\Sigma^3$ is locally isometric to $S^3$. Thus the only near horizon geometry preserving 8 supersymmetries with $h\not=0$  is
 $AdS_3 \times S^3$.

\newsection{Horizons with $h=0$}

\subsection{Geometry of $N=1$ horizons}

Let us now turn to horizons with $h=0$.  Clearly in such a case, the 3-form field strength
 of the gravitational multiplet vanishes $H=0$, and the near horizon geometry is a product $\bR^{1,1}\times {\cal S}$. It remains to determine the geometry of ${\cal S}$.

{}For this first observe that the tensor multiplet scalars are constant and the associated 3-form field strengths vanish. The proof for this is similar to that given for the horizons with $h\not=0$. In particular, it utilizes the field equations of the tensor multiplet scalars
as described in the equations (\ref{xx1}) and (\ref{xx2}), with $h=0$, and
the argument developed around (\ref{xu1}).

 The Einstein equation expresses the Ricci tensor $\tilde R$ of ${\cal S}$ in terms of the hypermultiplet
 scalars as
 \bea
\tilde R_{ij}
= 2g_{\ui\uj}\partial_i\phi^\ui\partial_j\phi^\uj~.
\la{xyz2}
\eea
 The latter are also restricted by the Killing spinor equations as in (\ref{1hyp1}). Observe
 that after an appropriate identification of frame directions of ${\cal S}$ with the matrices
 $(\tau^i)=( 1_{2\times 2}, i\sigma^{r'})$, (\ref{1hyp1}) can be written as
 \bea
 (\tau^i)^\uA{}_\uB\, \partial_i\phi^\ui\, E_\ui^{\ua\uB}=0~,
 \eea
or equivalently in terms of the quaternionic structure $I_{r'}$ of ${\cal Q}$ as
 \bea
 (K_{i})^\ui{}_\uj\, \partial^i\phi^\uj=0~,
 \la{1hypu}
 \eea
 where $(K_i)=(1_{4n_H\times 4n_H}, -I_{r'})$ and  $1_{4n_H\times 4n_H}$ is the identity tensor.

 If the hypermultiplet scalars are constant, then the rhs of (\ref{xyz2}) vanishes and ${\cal S}$ is a hyper-K\"ahler manifold. So
 it is locally isometric to either $K_3$ or $T^4$. As we shall see such horizons exhibit supersymmetry enhancement to at least $N=4$.

The existence of horizons with strictly $N=1$ supersymmetry
 depends on the existence of non-trivial solutions for (\ref{1hypu}) such that the rhs of
 (\ref{xyz2}) does not vanish. This in turn may depend
 on the choice of the 4-manifold ${\cal S}$ and that of the quaternionic K\"ahler manifold
 ${\cal Q}$. As a result, this is a rather involved question, and possibly model dependent, which  we shall not explore  further here.

 \subsection{Geometry of $N=2$ and $N=4$ horizons}

The second Killing spinor of $N=2$ horizons with  $h=0$  can be chosen as
\bea
\epsilon_2=i(1-e_{1234})~.
\eea
In such case, and in agreement with the general classification results of \cite{akyolgp},
${\cal S}$ is a K\"ahler manifold. In addition,  the hypermultiplet scalars are holomorphic maps
from ${\cal S}$ into the hypermultiplets Quaternionic K\"ahler manifold.  Again, the existence
of  such horizons with strictly 2 supersymmetries  depends on the existence of such non-trivial  holomorphic maps

The two remaining Killing spinors of $N=4$ horizons with $h=0$ can be chosen as
\bea
\epsilon_3=e_{12}-e_{34}~,~~~\epsilon_4=i(e_{12}+e_{34})~.
\eea
The general classification results of \cite{akyolgp} imply that the hypermultiplet scalars
are constant as a consequence of the hyperini KSEs. Therefore ${\cal S}$ is hyper-K\"ahler
and so locally isometric to either $K_3$ or $T^4$. In the latter case, there is supersymmetry enhancement to $N=8$.

So far we have considered Killing spinors annihilated by the lightcone projection operation
$\Gamma_+$. As a result,  they have a non-compact isotropy group in $Spin(5,1)\cdot Sp(1)$. We could demand
that the near horizon geometries $\bR^{1,1}\times {\cal S}$ admit Killing spinors with compact isotropy groups. In such case, the only solution is $\bR^{1,1}\times T^4$ which preserves 8 supersymmetries.
 Some of the results of this section are tabulated in table 2.

\begin{table}
\centering
\fontencoding{OML}\fontfamily{cmm}\fontseries{m}\fontshape{it}\selectfont
\begin{tabular}{|c|c|c|}\hline
$${N}$$& ${\rm hol}(\tilde\nabla)$ &{\rm Geometry} \quad {\rm of}\quad${\cal S}$
 \\
\hline\hline

$1$&$Sp(1)\cdot Sp(1)$&{\rm Riemann}
\\
\hline
$2$&$U(2)$&{\rm K\"ahler}
\\
\hline
$4$&$Sp(1)$&{\rm hyper-K\"ahler}
\\
\hline
\end{tabular}
\label{tab2}
\begin{caption}
{\small {\rm ~~Some geometric data of the horizon geometries with $h=0$ are described. In the first column, we give  the number of supersymmetries preserved. In the second column, we present  the holonomy groups
of the  Levi-Civita connection of ${\cal S}$, and in the third we give the geometry of ${\cal S}$.
}}
\end{caption}
\end{table}

\vskip 0.5cm
\vskip 0.5cm
{\bf Acknowledgments:}~We  thank Jan Gutowski and Fabio Riccioni
for many helpful
discussions.
 MA is supported by the STFC  studentship grant  ST/F00768/1. GP is
 partially supported
by the EPSRC grant EP/F069774/1 and the STFC rolling grant ST/G000/395/1.

\setcounter{section}{0}
\setcounter{subsection}{0}

\appendix{Supersymmetric horizons}

\subsection{Lightcone intergrability of KSEs}

The gravitino KSE is
\bea
{\cal D}_\mu \epsilon=0~,
\label{gkse1}
\eea
where
\bea
{\cal D}_\mu= \hat\nabla_\mu +\mathcal{C}_\mu^{r'}\rho_{r'}=
\partial_{\mu} + \frac{1}{4}\Omega_{\mu,\nu\rho}\Gamma^{\nu\rho} - \frac{1}{8}H_{\mu\nu\rho}\Gamma^{\nu\rho} +\mathcal{C}_\mu^{r'}\rho_{r'}~.
\label{gkse2}
\eea
We  also identify the stationary Killing vector field of the black hole $\partial_u$ with
the Killing vector constructed as Killing spinor bi-linear. Since the latter is null \cite{akyolgp},  $\Delta=0$ in
(\ref{bhdata}). As a result,
the non-vanishing components of the frame connection associated with the Levi-Civita connection of the spacetime are
\be
\Omega_{+,-i} &=& -\frac{1}{2}h_i~,~~~\Omega_{+,ij} = -\frac{1}{2}r(dh)_{ij}~,~~~\Omega_{-,+i} = -\frac{1}{2}h_i~,\nn
\cr
\Omega_{i,+-} &=& \frac{1}{2}h_i~,~~~\Omega_{i,+j} = -\frac{1}{2}r(dh)_{ij}~,~~~\Omega_{i,jk} = \tilde{\Omega}_{i,jk}~.
\label{spinconnection}
\ee
For later use, the anti-self duality of $H$  implies
\be
H_{+-i}&=&\frac{1}{3!}\epsilon_{+-ijkl}H^{jkl}~,\nn
&=&\frac{1}{3!}\epsilon_{+-ijkl}v_{\underline r}dW^{{\underline r}jkl}~.
\ee

The KSEs can be integrated along the light-cone directions for the fields given in
(\ref{bhdata}). For this, decompose
the Killing spinor $\epsilon$ as
\be
\epsilon=\epsilon_+ + \epsilon_-~,~~~~~~~\Gamma_{\pm}\epsilon_{\pm}=0~.
\ee
The $-$ component of the gavitino KSE, (\ref{gkse1}),  gives
\be
\partial_{-}\epsilon - \frac{1}{4}(v_\ur dS^\ur-N-(S+1)h)_i\Gamma^{i}\Gamma_-\epsilon_+ = 0~,
\ee
where we have used the expression for the frame connection stated above, the expression
for the fields in (\ref{bhdata}) and that $\mathcal{C}_-^{r'} =0$ as the scalars do not depend on $(u,r)$ in the near horizon limit.
We have also set $N=v_\ur N^\ur$ and $S=v_\ur S^\ur$. Noting that $\partial_-=\partial_r$ and $\partial_+=\partial_u$ and upon integration, we find
\be
\epsilon_+&=&\phi_+~,\nn
\epsilon_-&=&\phi_- + \frac{1}{4}r(v_{\underline r}dS^{\underline r}-N-(S+1)h)_i\Gamma^{i}\Gamma_-\phi_+~,
\label{phi}
\ee
where $\phi_{\pm}$ are independent of $r$.

Similarly, the $+$ component of the gravitino KSE gives
\bea
\partial_{+}\epsilon + \frac{1}{4}(v_{\underline r}dS^{\underline r}-N-(S-1)h)_i\Gamma^{i}\Gamma_+\epsilon_- - \frac{1}{8}r(h\wedge N-v_{\underline r}dN^{\underline r} -(S-1)dh)_{ij}\Gamma^{ij}\epsilon = 0~.
\eea
Substituting in (\ref{phi}) into the $+$ component of the gravitino KSE, we get
\bea
&&\partial_{+}\left(\phi + \frac{1}{4}r(v_{\underline r}dS^{\underline r}-N-(S+1)h)_i\Gamma^{i}\Gamma_-\phi_+ \right)\nn
&&+ \frac{1}{4}(v_{\underline r}dS^{\underline r}-N-(S-1)h)_i\Gamma^{i}\Gamma_+\left(\phi_- + \frac{1}{4}r(v_{\underline r}dS^{\underline r}-N-(S+1)h)_i\Gamma^{i}\Gamma_-\phi_+ \right)\nn
 &&- \frac{1}{8}r(h\wedge N-v_{\underline r}dN^{\underline r} -(S-1)dh)_{ij}\Gamma^{ij}\left(\phi + \frac{1}{4}r(v_{\underline r}dS^{\underline r}-N-(S+1)h_i\Gamma^i\Gamma_-\phi_+ \right)
 \cr
 &&~~~~~~~~~~~~~~~~~= 0~.
 \la{xxx3}
\eea
This equation is valid in every order in $r$. As a result  $\mathcal{O}(r^0)$ order term gives
\bea
\partial_+\phi +\frac{1}{4}(v_{\underline r}dS^{\underline r}-N-(S-1)h)_i\Gamma^{i}\Gamma_+\phi_-=0~,
\eea
which can be solved to find
\bea
\phi_+ &=& \eta_+ - \frac{1}{4}u(v_{\underline r}dS^{\underline r}-N-(S-1)h)_i\Gamma^{i}\Gamma_+\eta_- ~,\nn
\phi_- &=& \eta_{-}~,
\eea
where $\eta_{\pm}$ is independent of $r$ and $u$.
As a result, the components $\epsilon_\pm$ of the Killing spinor can be written in terms
of $\eta_\pm$ as
\bea
\epsilon_+ &=& \eta_+ - \frac{1}{4}u(v_{\underline r}dS^{\underline r}-N-(S-1)h)_i\Gamma^{i}\Gamma_+\eta_- ~,\nn
\epsilon_- &=& \eta_{-}+ \frac{1}{4}r(v_{\underline r}dS^{\underline r}-N-(S+1)h)_i\Gamma^{i}\Gamma_-\eta_+ \nn
&&+\frac{1}{8}ur(v_{\underline r}dS^{\underline r}-N-(S+1)h)_i(v_{\underline r}dS^{\underline r}-N-(S-1)h)_j\Gamma^{i}\Gamma^{j}\eta_-  ~.
\label{spinor}
\eea
The remaining conditions implied by (\ref{xxx3})  are algebraic which   will be considerably simplified after the analysis of the next section. These are
\bea
\alpha_i\beta_j\Gamma^i\Gamma^j\eta_+ + \gamma_{ij}\Gamma^{ij}\eta_+&=&0~, \label{c1}\\
\gamma_{ij}\beta_k\Gamma^{ij}\Gamma^k\eta_+&=&0~,\\
\beta_i\alpha_j\Gamma^i\Gamma^j\eta_- - \gamma_{ij}\Gamma^{ij}\eta_-&=&0~,\\
\alpha_i\beta_j\alpha_k\Gamma^i\Gamma^j\Gamma^k\eta_- + \gamma_{ij}\alpha_k\Gamma^{ij}\Gamma^k\eta_-&=&0~,\\
\gamma_{ij}\beta_k\alpha_l\Gamma^{ij}\Gamma^k\Gamma^l\eta_-&=&0~,\label{c5}
\eea
where
\bea
\alpha_i&=&(v_{\underline r}dS^{\underline r}-N-(S-1)h)_i~,\\
\beta_i&=&(v_{\underline r}dS^{\underline r}-N-(S+1)h)_i~,\\
\gamma_{ij}&=&(h\wedge N-v_{\underline r}dN^{\underline r} -(S-1)dh)_{ij}~.
\eea
These are in fact the same constraints as those found for the heterotic horizons in \cite{hh}.

\subsection{Stationary and spinor bi-linear vector fields}
Additional restrictions on $\eta_\pm$ can be derived for horizons preserving one supersymmetry arising from  the identification of stationary black hole Killing vector field $\partial_u$ with that constructed as a Killing spinor bi-linear.    This identification implies that the components of the  1-form associated with the latter are
\bea
X_+=0~,~~~X_-=1~,~~~X_i=0~.
\label{bilinearcomponents}
\eea

The $\eta_\pm$  spinors  can be expanded in   basis  of symplectic Majorana-Weyl spinors as
\bea
\eta_+&=&a_1(1+e_{1234})+a_2i(1-e_{1234})+a_3(e_{12}-e_{34})
+a_4i(e_{12}+e_{34})~,
\cr
\eta_-&=&b_1(e_{15}+e_{2345})+b_2i(e_{15}-e_{2345})+b_3(e_{25}-e_{1345})+
b_4i(e_{15}+e_{2345})~,
\eea
where all components depend on the coordinates $y$ of ${\cal S}$.
The field data (\ref{bhdata}) are covariant under local $Spin(4)\cdot Sp(1)$ gauge transformations of ${\cal S}$. So these   can be used  to choose $\eta_\pm$ as
\bea
\eta_+&=&a(y)(1+e_{1234})~,\cr
\eta_-&=&b(y)(e_{15}+e_{2345})~.
\eea
The next step is to consider the spinor bilinear
\bea
Y_Ae^A=<B\epsilon^*,\Gamma_A\epsilon>e^A~,
\eea
where $B=\Gamma_{06789}$. In order to satisfy the relations in (\ref{bilinearcomponents}), we require the + component for the spinor bilinear to vanish. This in particular means that  $Y_+|_{r=0}=0$, and as a consequence we find
\bea
\eta_-=0~.
\eea
Therefore we can write the spinor in (\ref{spinor}) as
\bea
\epsilon &=& \eta_+ + \frac{1}{4}r(v_{\underline r}dS^{\underline r}-N-(S+1)h)_i\Gamma^{i}\Gamma_-\eta_+
\eea
Since the bilinear components on the horizon are independent of r, the next requirement we impose is for the $\mathcal{O}(r)$ term in the bilinear to vanish. This means
\bea
<B(1+e_{1234}),\Gamma_A(v_{\underline r}dS^{\underline r}-N-(S+1)h)_i\Gamma^{i}\Gamma_-\eta_+>=0~,
\eea
from which we obtain the condition
\bea
v_{\underline r}dS^{\underline r}-N-(S+1)h=0~.
\label{thecondition}
\eea
This simplifies the Killing spinor as
\bea
\epsilon=\eta_+=a(x)(1+e_{1234})~.
\eea
Finally calculating $Y_-$ and comparing this to $X_-$, we find
\bea
-2\sqrt{2}a^2=1~,
\eea
i.e. $a$ is a constant, which without loss of generality can be set to 1. This means
\bea
\epsilon=1+e_{1234}~.
\eea
This choice of a Killing spinor for the horizon geometries is the same as that  for general solutions of the KSEs of 6-dimensional supergravity preserving one supersymmetry.  This will be used to simplify the analysis of near horizon geometries.

\subsubsection{Further analysis of the gravitino KSE}
Revisiting the $+$ component of the gravitino KSE for $\epsilon=1+e_{1234}$, one finds that
\bea
\left(h\wedge N-v_{\underline r}dN^{\underline r} -(S-1)dh\right)_{ij}\Gamma^{ij}\eta_+ = 0~.
\eea
As a consequence  all algebraic  conditions (\ref{c1}-\ref{c5}) are also satisfied.

Next consider the  i-component of the gravitino KSE. After separating the various orders in $r$, one finds that
\bea
\mathcal{\tilde{D}}_i\eta_+=\left(\partial_i + \frac{1}{4}\tilde{\Omega}_{i,jk}\Gamma^{jk}-\frac{1}{8}v_{\underline r}(dW^{\underline r})_{ijk}\Gamma^{jk}+\mathcal{C}^{r'}_i\rho_{r'}\right)\eta_+=0~.
\eea
and
\bea
(h\wedge N -v_{\underline r}dN^{\underline r} -(S+1)dh)_{ij}\Gamma^{j}\eta_+=0~.
\label{condition2}
\eea
Using $\eta_+=1+e_{1234}$ in the last equation, we find that
\bea
h\wedge N -v_{\underline r}dN^{\underline r} -(S+1)dh=0~.
\eea
As a result, the 3-form $H$ simplifies as
\bea
H={\bf e}^+ \wedge {\bf e}^- \wedge h+r{\bf e}^+ \wedge dh+v_{\underline r}dW^{\underline r}~.
\eea

\subsection{Tensorini KSEs}
Now consider the tensorini KSEs
\bea
\left(T_{\mu}^{\underline{M}}\Gamma^{\mu}-\frac{1}{12}H_{\mu\nu\rho}^{\underline{M}}\Gamma^{\mu\nu\rho}\right)\epsilon=0\ ,
\eea
where $\epsilon=\eta_+=1+e_{1234}$.
This has been solved in \cite{akyolgp} where we have found
\bea
T_{+}^{\underline{M}}=0,\qquad H_{+\alpha}^{\underline{M}\ \alpha}=H_{+\alpha\beta}^{\underline{M}} & = & 0,\nn
T_{\overline{\alpha}}^{\underline{M}}-\frac{1}{2}H_{-+\overline{\alpha}}^{\underline{M}}-\frac{1}{2}H_{\overline{\alpha}\beta}^{\underline{M}\ \beta} & = & 0.
\eea
These together with the self-duality of $H^\uM$  imply that
\bea
H_{+ij}^{\underline M}=0~,~~~
T_i^{\underline M} = H_{-+i}^{\underline M} ~.
\eea
Comparing this with the expression of $H^\uM$ in (\ref{bhdata}), we get the expression for $H^\uM$ in (\ref{n1fields}). In particular, we have
\bea
x^\uM_\ur dS^\ur-N^\uM-S^\uM h=-T^\uM~,~~~h \wedge N^\uM - x^\uM_\ur dN^\ur - S^\uM dh =0~,
\eea
where $N^\uM=x^\uM_\ur N^\ur$ and $S^\uM=x^\uM_\ur S^\ur$.

\subsection{Hyperini KSEs}
A direct application of the results in \cite{akyolgp} and using that
the scalars of the hyper-multiplet do not depend on the coordinates $(u,r)$
reveal that the hyperini KSEs imply that
\bea
-V_{1}^{\underline{a1}}+V_{\overline{2}}^{\underline{a2}}=0~,~~~ V_{2}^{\underline{a1}}+V_{\overline{1}}^{\underline{a2}}=0~,
\eea
for $\epsilon=1+e_{1234}$.


\end{document}